\documentclass{epl}

\title{Does a Brownian particle equilibrate?}
\author{A. V. Plyukhin}
\institute{
  Department of Physics and Engineering Physics,
University of Saskatchewan\\Saskatoon, SK S7N 5E2, Canada
}


\begin{document}

\maketitle

\begin{abstract}
The conventional equations of Brownian motion can be derived from
the first principles to order $\lambda^2=m/M$,
where $m$ and $M$ are the masses
of a bath molecule and a Brownian particle respectively.
We discuss the extension to order $\lambda^4$
using 
a perturbation analysis of the Kramers-Moyal expansion. 
For the momentum distribution such  method
yields  an equation whose stationary solution is
inconsistent  with Boltzmann-Gibbs statistics.
This property originates entirely from non-Markovian corrections
which are negligible in lowest order but contribute to order
$\lambda^4$.
\end{abstract}

Dynamical justification of equilibrium 
statistical mechanics is  a long-standing problem
which can be traced to Einstein's criticism of the 
statistical definition of probability of macrostates~\cite{Cohen1}. 
The renewed interest stems primarily from  
recent development in the theory of  nonextensive
systems for which  the validity of classical
Boltzmann-Gibbs (BG) statistics is not obvious.  
A number of dynamical models generating non-canonical distributions 
have been suggested lately~\cite{Beck},  
but for a truly conservative Hamiltonian system deviations 
from the BG statistics 
have never been dynamically justified.
On the other hand, the BG statistics itself has been recovered 
directly from dynamics only for a few simplified models.
Even for a dilute gas  
it has been done only in the limit of  pair collisions.  
The {\it H}-theorem can be readily proved when dynamics is 
described by the master equation with transition rates
obeying  the detailed balance condition~\cite{VK}.  
The latter can be deduced  from reversibility of the microscopic processes
involved. However, the master equation is usually 
a Markovian approximation of an originally non-Markovian exact
equation for a targeted variable.
The detailed balance arguments do not necessarily apply
in this case, as will be discussed at the end of the Letter.

A phenomenon where  relaxation  to equilibrium 
may be traced, under reasonable assumptions, from the first principles is
Brownian motion of a heavy particle (B particle) of mass $M$ 
in an infinite bath of light molecules of mass $m\ll M$.
Using an appropriate projection operator technique  one can 
get to order $\lambda^2=m/M$ the Langevin equation (LE) 
and  the corresponding 
Fokker-Planck equation 
which describe exponential
relaxation of the particle's momentum to the Maxwellian distribution.  
However, a generalization to higher orders in $\lambda$ is not trivial
and has been studied in detail only for the very special 
Rayleigh model where 
the B-particle interacts with ideal gas molecules via the hard-wall potential
and the molecules do not interact with each other at all~\cite{VK,PhysA}.

In this Letter we discuss
Brownian motion to order $\lambda^4$ in the general
case of continuous interaction. The essential difference with 
the Rayleigh model is that the Langevin equation in general
is non-Markovian. However,
one may perform systematic expansion around the Markovian limit 
converting the LE into a 
local equation with appropriate corrections.
To order $\lambda^2$ these non-Markovian corrections are  negligible, 
but must be retained to order $\lambda^4$.
The main result of the Letter is that  non-Markovian  corrections
generate additional terms in the $\lambda^4$-order Fokker-Planck
equation 
which make it inconsistent
with the BG statistics.

Let us first outline the elements of the dynamical theory of
Brownian motion~\cite{MO,PS}. The basic   
assumption is that
not too far from equilibrium 
the relation of 
the particle momentum $P$ and typical momentum  of a bath molecule
$p_m$ is given by
the equipartition  theorem,  $p_m/P\sim\lambda=\sqrt{m/M}$, and therefore
the scaled momentum of the  B-particle $p=\lambda P$ is of order $\lambda^0$. 
It is assumed that writing the equation of motion for the scaled momentum
$p$ rather than for the true momentum $P$ enables one
to extract the overall dependence of the problem on the small parameter
$\lambda$.
Then starting with the exact equation of motion for the B-particle 
and using an appropriate projection operator and
perturbation techniques
one can obtain to order $\lambda^2$
the non-Markovian LE for the particle's  scaled
momentum 
\begin{eqnarray}
\frac{dp(t)}{dt}=-\lambda^2\,\int_0^t d\tau M_0(\tau)p(t-\tau)+\lambda F(t).
\label{NMLE}
\end{eqnarray}
Here the  ``random'' force $F(t)$ depends on 
the coordinates and momenta of bath molecules
and fluctuates rapidly with mean zero.
It is assumed that $F$  may be expanded in powers of $\lambda$ 
\begin{eqnarray}
F(t)=F_0(t)+\lambda F_1(t)+\lambda^2 F_2(t)+\cdots
\label{F_expansion}
\end{eqnarray}
where the first term $F_0(t)$ has a  meaning of the force exerted by
the bath molecules on the B-particle fixed in space. In other words,
$F_0(t)$ is a fluctuating pressure on the infinitely heavy particle. 
Only this term should be taken
into account in the $\lambda^2$-order equation,   
and the fluctuation-dissipation theorem has the form
\begin{eqnarray}
M_0(t) =(\beta/m)\langle F_0(0)F_0(t)\rangle,
\label{M0}
\end{eqnarray}
where $\beta$ is the inverse temperature of the bath,
and the average is taken with respect to canonical 
distribution of the bath in the field of the B-particle
held fixed.
Next, noting that 
$
p(t-\tau)=p(t)-\int_{t-\tau}^t dt'\dot{p}(t')=p(t)+O(\lambda)
$
one finds that non-Markovian effects are not important at order  
$\lambda^2$, and therefore the LE (\ref{NMLE}) can be written in
the conventional local form
\begin{eqnarray}
\frac{dp(t)}{dt}=-\lambda^2\,\gamma_0\, p(t)+\lambda F_0(t),
\label{LE}
\end{eqnarray}
with $\gamma_0=\int_0^\infty dt M_0(t)$.
Integrating this equation
one finds that $\langle
p^2(t)\rangle$ relaxes  exponentially to
the equilibrium value $m/\beta$ in
agreement with  BG statistics. 

Let us also remind the reader how  the Fokker-Planck equation 
(FPE) for the distribution 
function $f(p,t)$, corresponding  to the LE (\ref{LE}), may be derived.  
Starting with a  generic master equation, which is 
valid for an arbitrary Markovian process, one can follow the familiar 
procedure~\cite{VK} 
to transform  it into the Kramers-Moyal expansion
\begin{eqnarray}
\frac{\partial f(p,t)}{\partial t}=
\sum_{n=1}\frac{1}{n!}\left(-\frac{\partial}{\partial p}\right)^n
\Bigl\{\alpha_n(p)f(p,t)\Bigr\}
\label{KME}
\end{eqnarray}
where the coefficients are
\begin{eqnarray}
\alpha_n(p)=\lim_{\tau\to 0}\frac{1}{\tau}
\langle [p(t+\tau)-p(t)]^n\rangle.
\label{as}
\end{eqnarray} 
These coefficients may be evaluated integrating the LE (\ref{LE})
for small $\tau$,
\begin{eqnarray}
&&p(t+\tau)-p(t)= -\lambda^2\,\gamma_0 \, p(t)\, \tau+
\lambda\int_t^{t+\tau}dt' F_0(t').
\label{aux1}
\end{eqnarray} 
This formula corresponds to a coarse-grained
description with a time resolution much shorter than the relaxation
time of the particle's momentum $\tau_p$ and much longer than the correlation
time for the random force $\tau_c$.
One can not require 
a finer time resolution because
the LE itself in the form
(\ref{LE}) is valid only on a time scale $t\gg\tau_c$. 
Two remarks are in order here. First, 
the route we follow is not a good one when $\tau_c$ does not exist, for
instance when $C(t)=\langle F_0(0)F_0(t)\rangle\sim t^{-\sigma}$. 
We ignore the slow hydrodynamic modes of the bath responsible for
long-time tails of correlation functions.
Second,
the coarse-grained description implies that 
the moments
$\langle [p(t+\tau)-p(t)]^n\rangle$ in Eq.(\ref{as})
 must be calculated
in the limit $\tau\gg \tau_c$. Only after that must the operation
$\lim_{\tau\to 0}\frac{1}{\tau}(\cdots)$  be taken.

In general the ``random'' force in the LE is neither delta-correlated
nor Gaussian, and therefore the Kramers-Moyal  expansion is not
truncated.  However, only the two first terms involve 
contributions of order $\lambda^2$. Then to order $\lambda^2$ 
the expansion  turns into the familiar second-order
Fokker-Planck equation
\begin{eqnarray}
\frac{\partial f(p,t)}{\partial t}=\lambda^2 D_2\,f(p,t),
\,\,\,\,\,\,\,\,\,
\label{FPE}
D_2=a_1\,\frac{\partial}{\partial p}\,p+
a_2\,\frac{\partial^2}{\partial p^2}.
\end{eqnarray}
Since the pressure $F_0(t)$ is a stationary process,
the coefficients have the form
\begin{eqnarray}
&&a_1=\gamma_0=\frac{\beta}{m}
\int_0^\infty dt \,\langle F_0(0)F_0(t)\rangle,\,\,\,\,\,
a_2=\frac{1}{2\tau}\,\int_0^\tau dt_1\int_0^\tau dt_2\,
\langle F_0(t_1)F_0(t_2)\rangle.
\end{eqnarray}
Recall  that the double integral in 
the last expression must be taken in the limit $\tau\gg \tau_c$.
The subsequent limit $\tau\to 0$ is not necessary because
the integral  grows 
linearly with time for $\tau\gg \tau_c$. Indeed,  
taking into account that $\langle F_0(t_1)F_0(t_2)\rangle=
C(|t_1-t_2|)$ one obtains
\begin{eqnarray}
\int_0^\tau \!\!dt_1\!\!\int_0^\tau\! dt_2\langle F_0(t_1)F_0(t_2)\rangle=
2\int_0^\tau \!dt_1\!\int_0^{t_1} \!dt_2C(t_1-t_2)=
2\int_0^\tau \!dt_1 (\tau-t_1)C(t_1).
\end{eqnarray} 
For $\tau\gg \tau_c$ this integral equals 
$2\tau\,\int_0^\infty \!dt_1 C(t_1)$.
With this result, one finds the relation
\begin{eqnarray}
a_1=\frac{\beta}{m}\,a_2.
\label{condition1}
\end{eqnarray}
This means that 
the stationary solution of the  equation (\ref{FPE}), 
$f(p)=c\exp\left(-\frac{a_1}{2a_2}p^2\right)$,
is  Maxwellian.

Our aim is to generalize the above approach  to order $\lambda^4$.
(The $\lambda^3$-order damping  force vanishes due to symmetry for
a homogeneous bath.) The
$\lambda^4$-order LE has the form
\begin{eqnarray}
\frac{dp(t)}{dt}= -
{\lambda}^2\int_0^t d\tau\,\,
M_1(\tau) p(t-\tau)-{\lambda}^4\int_0^t d\tau\,\,
M_2(\tau) p^3(t-\tau)+\lambda F(t).
\label{NMNLE}
\end{eqnarray}  
A microscopic derivation of this equation was recently discussed 
in detail in~\cite{PS}. 
It differs from the $\lambda^2$-order LE (\ref{NMLE}) not only
by the presence of the nonlinear damping term involving $p^3$, but also 
by additional corrections of order $\lambda^2$ to 
the  memory kernel for linear damping
\begin{eqnarray}
M_1(t)=M_0(t)+\lambda^2\,\delta M(t).
\label{M1}
\end{eqnarray}
Another difference is that the ``random'' force in Eq. (\ref{NMNLE}) can not 
be approximated by the pressure term $F_0(t)$, as 
in the  linear LE (\ref{LE}), but must involve the  
corrections of higher orders
$F_1$ and $F_2$, see Eq.(\ref{F_expansion}).
The correction to the linear damping $\delta M$ and the  
nonlinear damping kernel $M_2(t)$ can be expressed in terms of 
correlation functions involving $F_0$, $F_1$ and $F_2$ \cite{PS},
but their explicit forms are not important for our purpose
here.

Our first step is to find a Markovian approximation for the non-Markovian 
equation (\ref{NMNLE}) using the same trick as for the linear LE, i.e.
writing
$\phi(t-\tau)=\phi(t)-\int_{t-\tau}^t dt'\dot{\phi}(t')$ for
$\phi(t)=p^n(t)$.
Since $\frac{d}{dt} p^n(t)\sim\dot p(t)=O(\lambda)$ according 
to (\ref{NMNLE}), 
the non-Markovian correction for the nonlinear damping is
of order $\lambda^5$ and should be neglected,
\begin{eqnarray}
&&-{\lambda}^4\int_0^t d\tau\,\,
M_2(\tau) p^3(t-\tau)\to
-{\lambda}^4\,p^3(t)\int_0^\infty d\tau\,\,
M_2(\tau)
\end{eqnarray}
However, for the linear damping term the non-Markovian corrections  
are  of order $\lambda^3$ and $\lambda^4$ and must be retained. 
Indeed, let us write the linear damping term  
as a local expression plus a correction $\Delta(t)$,
\begin{eqnarray}
-{\lambda}^2\int_0^t d\tau
M_1(\tau)p(t-\tau)=
-{\lambda}^2p(t)\int_0^t d\tau
M_1(\tau)+\Delta(t).
\end{eqnarray}
The correction 
\begin{eqnarray}
&&\Delta(t)={\lambda}^2\,\int_0^t d\tau\,\,
M_1(\tau)\int_{t-\tau}^t\,dt'\,\dot{p}(t')
\end{eqnarray}  
can be evaluated estimating $\dot{p}(t)$ from the LE (\ref{NMNLE}).
For $t\gg\tau_c$ one obtains to order $\lambda^4$
\begin{eqnarray}
&&\!\!\!\!\!\Delta(t)=\lambda^3\,\int_0^t\!\! d\tau\,\,
M_0(\tau)\int_{t-\tau}^t\,dt'
 F_0(t')
-\lambda^4\,p(t)\int_0^{\infty}\!\!dt\, M_0(t)\,\int_0^{\infty}\!\! dt\,
M_0(t)\,t.
\end{eqnarray}  
We shall write this expression in the form
\begin{eqnarray}
\Delta(t)=
\lambda^3\,F^\star(t)-\lambda^4\,\gamma^\star\,p(t).
\label{delta}
\end{eqnarray}    
Here the first term involves a rapidly changing function
\begin{eqnarray}
F^\star(t)=\frac{\beta}{m}\,\int_0^t d\tau\,\,
C(\tau)\int_{t-\tau}^t\,dt'
 F_0(t')
\label{Fstar}
\end{eqnarray}  
and may be considered as an additional  contribution
to the random force. Note that $F^\star(t)$ is 
approximately stationary on the time scale $t\gg \tau_c$. 
The second term  in Eq.(\ref{delta}) with
\begin{eqnarray}
\gamma^\star=\left(\frac{\beta}{m}\right)^2\int_0^\infty dt\,C(t)\,
\int_0^\infty dt\,C(t)\,t
\label{gstar}
\end{eqnarray} 
can be interpreted as an additional contribution
to the linear damping.

As a result, 
the LE (\ref{NMNLE}) can be written in  local form as follows
\begin{eqnarray}
\frac{dp(t)}{dt}= 
-\lambda^2\gamma_1\,p(t)-{\lambda}^4\gamma_2\,p^3(t)+\lambda\, \xi(t).
\label{NLE}
\end{eqnarray}  
Here $\xi(t)$ is the  ``random'' force  with a $\lambda^2$-order 
non-Markovian correction 
\begin{eqnarray}
\xi(t)&=&F(t)+\lambda^2 F^\star(t)=
F_0(t)+\lambda F_1(t)+\lambda^2 F_2(t)+\lambda^2 F^\star(t).
\label{force_expansion}
\end{eqnarray}
The linear damping coefficient in (\ref{NLE}) takes the form
\begin{eqnarray}
\gamma_1=\gamma_{0}+\lambda^2\,\delta\gamma+\lambda^2\gamma^\star,
\end{eqnarray}
with $\delta\gamma=\int_0^\infty dt\, \delta M(t)$, and the nonlinear
damping coefficient is $\gamma_2=\int_0^\infty dt \,M_2(t)$.

Integrating the LE (\ref{NLE}) for $\tau_c\ll\tau\ll\tau_p$ one gets
\begin{eqnarray}
p(t+\tau)\!-p(t)=
-\left[\lambda^2\,\gamma_1\, p(t)+\lambda^4\gamma_2\, p^3(t)\right]\,\tau+
\lambda\!\int_t^{t+\tau}\!\!dt'\xi(t').
\end{eqnarray} 
Then using Eq.(\ref{as}) one finds the coefficients $\alpha_n$ 
in the Kramers-Moyal
expansion (\ref{KME}),
\begin{eqnarray}
\alpha_1=-\lambda^2\gamma_1\,p-{\lambda}^4\gamma_2\,p^3
=-\lambda^2\gamma_0\,p-\lambda^4\,\gamma^\star\, p-\lambda^4\,
(\delta\gamma\,\, p+\gamma_2\, p^3)
\end{eqnarray}
and, recalling that $\xi(t)$ is stationary for $t\gg\tau_c$,  
\begin{eqnarray}
\alpha_n=\lambda^n
\lim_{\tau\to 0}\,\frac{1}{\tau}\int_0^\tau dt_1\int_0^\tau dt_2...
\int_0^\tau dt_n
\langle\xi(t_1)\xi(t_2)...\xi(t_n)\rangle
\label{cn}
\end{eqnarray}
for $n=2,3,4$. (The terms with $n>4$ do not contribute  
at order $\lambda^4$.)
In the above expression the integrals must be taken in the limit
$\tau\gg\tau_c$. 

According to (\ref{cn}), to get $\alpha_2$ to
order $\lambda^4$ one needs the correlation $\langle
\xi(t_1)\xi(t_2)\rangle$ to order $\lambda^2$,
\begin{eqnarray}
\lefteqn{\langle \xi(t_1)\xi(t_2)\rangle=
\langle F_0(t_1)F_0(t_2)\rangle+\lambda^2\langle F_0(t_1)F^\star(t_2)\rangle+
\lambda^2\langle F_0(t_2)F^\star(t_1)\rangle}\nonumber\\
&&+\lambda^2\langle
F_1(t_1)F_1(t_2)\rangle+\lambda^2\langle F_0(t_1)F_2(t_2)\rangle+
\lambda^2\langle F_0(t_2)F_2(t_1)\rangle.\nonumber
\end{eqnarray}
Similarly, $\alpha_3$ and $\alpha_4$ require the correlations 
to order $\lambda^1$ and $\lambda^0$, respectively:  
\begin{eqnarray}
&&\langle \xi(t_1)\xi(t_2)\xi(t_3)\rangle=
\lambda \langle F_0(t_1)F_0(t_2)F_1(t_3)\rangle
+\lambda \langle F_0(t_1)F_1(t_2)F_0(t_3)\rangle+
\lambda \langle F_1(t_1)F_0(t_2)F_0(t_3)\rangle,\nonumber\\
&&\qquad\qquad\qquad\qquad
\langle \xi(t_1)\xi(t_2)\xi(t_3)\xi(t_4)\rangle=
\langle F_0(t_1)F_0(t_2)F_0(t_3)F_0(t_4)\rangle.\nonumber
\end{eqnarray}

Using these formulas and  collecting the terms of the same order,  
one obtains the $\lambda^4$-order FPE in the form
\begin{eqnarray}
\frac{\partial f(p,t)}{\partial t}=
\Bigl\{\lambda^2 D_2+\lambda^4 D^\star_2 +
\lambda^4 D_4\Bigr\} f(p,t).
\label{NFPE}
\end{eqnarray}
Here the differential operator $D_2$ is the same as in the
$\lambda^2$-order
FPE (\ref{FPE}). The operator $D_2^\star$ originates 
from non-Markovian corrections 
and has the same structure as $D_2$,
\begin{eqnarray}
&&D_2^\star=b_1\,\frac{\partial}{\partial p}\,p+
b_2\,\frac{\partial^2}{\partial p^2}
\label{Dstar}
\end{eqnarray}
with $b_1=\gamma^\star$ given by Eq.(\ref{gstar}) and
\begin{eqnarray}
&&\!\!\!\!\!\!\!\!\!\!\!\!\!\!\!\!\!
b_2\!=\!\frac{1}{2\tau}\!\int_0^\tau\!\!\!\int_0^\tau\!dt_1dt_2
\Bigl\{\langle F^\star(t_1)F_0(t_2)\rangle
\!+\!\langle F^\star(t_2)F_0(t_1)\rangle\Bigr\}
\!=\!\frac{1}{\tau}\!\int_0^\tau\!\!\!\int_0^\tau\! dt_1\,dt_2
\langle F^\star(t_2)F_0(t_1)\rangle.
\end{eqnarray}

The operator $D_4$ in Eq.(\ref{NFPE})
is a differential operator of order four. 
It absorbs the terms of order $\lambda^4$ which are expressed in terms of 
time correlations $\langle F_0F_2\rangle$,
$\langle F_1F_1\rangle$, $\langle
F_0F_0F_1\rangle$, and 
$\langle F_0F_0F_0F_0\rangle$.
Since $\langle F_i(t)\rangle=0$,
the first three of these correlations  are actually
cumulants, which we denote by 
$\langle\!\langle \cdots\rangle\!\rangle$,
\begin{eqnarray}
&&\qquad\qquad\qquad\qquad\qquad
\langle F_iF_j\rangle\equiv
\langle F_i\rangle \langle F_j\rangle+
\langle\!\langle F_iF_j\rangle\!\rangle=
\langle\!\langle F_iF_j\rangle\!\rangle,\nonumber\\
&&\!\!\!\!\!\!\!\langle F_iF_jF_k\rangle\!\equiv\!
\langle F_i\rangle \langle F_j\rangle\langle F_k\rangle\!+\!
\langle F_i\rangle \langle\!\langle F_jF_k\rangle\!\rangle\!+\!
\langle F_j\rangle\langle\!\langle F_iF_k\rangle\!\rangle\!+\!
\langle F_k\rangle\langle\!\langle F_iF_j\rangle\!\rangle\!+\!
\langle\!\langle F_iF_jF_k\rangle\!\rangle
\!=\!\langle\!\langle F_iF_jF_k\rangle\!\rangle.
\nonumber
\end{eqnarray}
The cumulant expansion of the correlation
$\langle F_0F_0F_0F_0\rangle$,
which determines $\alpha_4$, 
involves the product of cumulants
$\langle\!\langle  F_0(t_1)F_0(t_2)\rangle\!\rangle 
\langle\!\langle F_0(t_3)F_0(t_4)\rangle\!\rangle $
and two similar terms. However, since  a product of cumulants 
depends only on two time differences, 
the corresponding contributions to 
the four-dimensional time integral in the expression for 
$\alpha_4$ vanish in the limit $\tau\to 0$.
Therefore only the cumulant
$\langle\!\langle F_0(t_1)F_0(t_2)F_0(t_3)F_0(t_4)\rangle\!\rangle$
contributes the expression for $\alpha_4$.

The fact that $D_4$ is linear in cumulants of $F_i$
means that $D_4$ is a linear functional of cumulants for
the density of bath particles $N(z,t)=\sum_i\delta(z-z_i(t))$,
where $z_i$ denotes
the coordinate-momentum pair $(x_i,p_i)$ of a bath particle.
In turn, one can observe that cumulants 
$\langle\!\langle N(z_1,t_1)N(z_2,t_2)...N(z_k,t_k)\rangle\!\rangle$
of any order $k$
depend  linearly on the concentration of bath molecules $n$.  
For instance,
in the expression for the product
$N(z_1,t_1)N(z_2,t_2)$ one can write the double sum as 
$\sum_{i,j}=\sum_{i\ne j}+\sum_{i=j}$ which gives
$\langle N(z_1,t_1)N(z_2,t_2)\rangle=
\langle N(z_1,t_1)\rangle\langle N(z_2,t_2)\rangle+
\sum_{i}\langle\delta(z_1-z_i(t_1))\delta(z_2-z_i(t_2))\rangle.
$
Here the second term on the right side 
is by definition the cumulant
$\langle\!\langle N(z_1,t_1)N(z_2,t_2)\rangle\!\rangle$
which is  obviously linear in $n$.

It follows from the above discussion that the operator
$D_4$ depends linearly on the concentration of bath molecules
$n$. In contrast, the first and second terms in 
$D_2^\star$
are  proportional respectively  to $\gamma^\star$ and 
the correlation $\langle F_0(t_1)F^\star(t_2)\rangle$,
which are both quadratic in $n$. 
The Maxwellian distribution
$f_M(p)\sim\exp\left(-\beta p^2/2m\right)$
 depends neither on $\lambda$ nor on $n$. 
Therefore, if $f_M(p)$ is a stationary solution of the FPE
(\ref{NFPE}), it must satisfy each term separately:
\begin{eqnarray}
D_2 f_M(p)=0,\,\,\,\,\,\,\,D_2^\star f_M(p)=0,\,\,\,\,\,\,\,D_4 f_M(p)=0.
\end{eqnarray}
The first relation, as was discussed at the beginning of the Letter, 
is satisfied. 
The validity of the relation 
$D_4 f_M=0$ is not our concern here. One can show 
that $D_4$ has the same structure as for the
Rayleigh model~\cite{VK,PhysA} and the validity of the relation 
$D_4 f_M=0$ can be explicitly checked  for the extended Rayleigh 
model~\cite{PS}. 
The central result of this Letter is that the second equation
$D_2^\star f_M=0$ is not satisfied. Indeed, this equation
implies that $b_1=\frac{\beta}{m}\,b_2$,
similar to the condition (\ref{condition1}) for $D_2$. 
Instead,  the relation with the additional
factor $1/2$ actually holds:  
\begin{eqnarray}
b_1=\frac{\beta}{2m}\, b_2.
\label{result}
\end{eqnarray}
Explicitly this relation has the form 
\begin{eqnarray}
\!\!\int_0^\infty dt C(t)\int_0^\infty dt C(t)t=
\frac{1}{2\tau}\int_0^\tau\!\int_0^\tau dt_1dt_2C^\star(t_1,t_2),
\label{result_ex}
\end{eqnarray}
where $C(t)=\langle F_0(0)F_0(t)\rangle$ and 
$C^\star(t_1,t_2)=(\beta/m)^{-1}
\langle F_0(t_1)F^\star(t_2)\rangle$.  
According to  (\ref{Fstar}), the latter correlation is given by  
\begin{eqnarray}
&&C^\star(t_1,t_2)=
\int_0^{t_2}dt'\,C(t')\int_{t_2-t'}^{t_2} dt''\,
C(|t_1-t''|).
\end{eqnarray}
Note that Eq.(\ref{result_ex}) is an asymptotic relation:
the double integral of $C^\star$ must be taken in the limit
$\tau\gg \tau_c$. The validity of the relation (\ref{result_ex}) 
may be verified directly, for instance, for 
$C(t)\sim e^{-t/\tau_c}$ or
for a Gaussian correlation.
A general proof of (\ref{result_ex})  
is rather lengthy and will be
presented elsewhere.
 
Since $D_2^\star f_M(p)\ne 0$, the stationary solution of the FPE (\ref{NFPE})
is not Maxwellian and in general can not be analyzed 
without identifying an explicit form of the operator $D_4$ which is
beyond our scope here. Suppose
however that the concentration of bath molecules $n$ is in a sense large and 
$D_4\sim n$ can be neglected compared to $D_2^\star\sim
n^2$. The $\lambda^4$-order FPE with $D_4$ dropped
is an  equation of second
order which has  the same structure as 
the $\lambda^2$-order equation (\ref{FPE}) but with modified
coefficients
$a^\star_i=a_i+\lambda^2 b_i$. Its stationary solution
$\exp\left(-\frac{a^\star_1}{2a^\star_2}p^2\right)$
is the Maxwellian distribution
$\exp\left(-\frac{\beta^\star}{2m}p^2\right)$
but with the inverse temperature
$\beta^\star=ma^\star_1/a^\star_2$ slightly lower than that for the
bath, $\beta$.
Taking into account relations (\ref{condition1}) and (\ref{result}),
one obtains $\beta^\star=\beta\,(1-\lambda^2\,b_1/a_1)$,
or 
\begin{eqnarray} 
\beta^\star=
\beta\left\{1-\lambda^2\,\frac{\beta}{m}\,\int_0^\infty dt\,C(t)\,t
\right\}.
\label{beta}
\end{eqnarray} 
This  result may be interpreted alternatively 
as a renormalization of 
the mass of the B particle 
$M\to M^\star=(\beta/\beta^\star) M$.

In conclusion, in this Letter we discussed the derivation of
the Fokker-Planck equation corresponding to 
the microscopic Langevin equation of order $\lambda^4$.
The procedure includes  the expansion of the  originally non-Markovian 
Langevin equation around the Markovian limit 
and evaluation  of the coefficients in the 
Kramers-Moyal expansion of the corresponding master equation. 
Let us  emphasize  that the method we used here is not just a truncation 
of the Kramers-Moyal expansion, which is known to yield an incorrect
description. 
Rather, we  transformed the Kramers-Moyal expansion into
the expansion in powers of small parameter $\lambda$
and applied a perturbation technique to obtain an equation to 
given order in $\lambda$. This  may be seen as a generalization 
of van Kampen's $1/\Omega$-expansion~\cite{VK,PhysA}. 
It is  found that to order $\lambda^4$ the method leads to the 
equation which has a non-Maxwellian stationary solution.
This feature originates 
entirely from non-Markovian corrections
collected in the operator $D_2^\star$, Eq.(\ref{Dstar}).
Since the method yields  correct
results for the Markovian Rayleigh model~\cite{PhysA}
at least to order $\lambda^4$, 
one may conclude that
either the  deviation from Maxwellian distribution
for conservative Brownian motion is a small but real effect, or
the procedure adopted here does not treat non-Markovian effects
properly. Note that we assumed but did not prove that
each term in $\lambda$-expansions is bounded for all $t$.
If the assumption is correct, 
the results of the Letter would mean that the conventional 
time-reversal arguments~\cite{VK},
leading to the detailed balance and, eventually, 
to the {\it H}-theorem, 
do not apply to Brownian motion
beyond the lowest order in $\lambda$.
This is perhaps not such a surprise recalling that
the right part of a Langevin equation is 
not a true force on the particle but rather an approximation 
to given order in $\lambda$.
Whether or not this truncated force corresponds to any effective Hamiltonian
is not known a priori, and therefore
the invariance for time reversal can not be invoked.
On the other hand, an {\it exact} closed equation for a targeted variable
is inevitably non-Markovian, and the standard
detailed balance arguments can not be applied either. 
As a referee noted, there is a similarity
with the Onsager relations
which follow from microscopic reversibility but 
become only
approximately valid when higher order corrections are included
in a fast variables elimination scheme~\cite{slip2}.




\end{document}